\documentclass[fleqn,multphys,vecphys]{svmult}

\usepackage{makeidx}   
\usepackage{graphicx}  
\makeindex             

\newcommand{\iDev}[1]{#1}
\newcommand{\iName}[1]{#1}
\newcommand{\iStreet}[1]{#1}
\newcommand{\iPostcode}[1]{#1}
\newcommand{\iCity}[1]{#1}
\newcommand{\iCountry}[1]{#1}
\newcommand{\email}[1]{\texttt{#1}}

\begin{document}

\title*{
Optical microcavities as quantum-chaotic model systems: Openness makes the difference!}
\toctitle{Optical microcavities as quantum-chaotic model systems: Openness makes the difference!}        
\titlerunning{Optical microcavities as quantum-chaotic model systems: Openness...}                                

\author{Martina Hentschel} 
\authorrunning{Martina Hentschel} 

\institute{\iDev{ }    
\iName{Max-Planck-Institut f\"ur Physik komplexer Systeme}, \newline
\iStreet{N\"othnitzer Str. 38},
\iPostcode{01187}
\iCity{Dresden},
\iCountry{Germany}\newline
\email{martina@pks.mpg.de}
}

\maketitle

\begin{abstract}
 Optical microcavities are open billiards for light in which electromagnetic waves can, however, be confined by total internal reflection at dielectric boundaries. These resonators enrich the class of model systems in the field of quantum chaos and are an ideal testing ground for the correspondence of ray and wave dynamics that, typically, is taken for granted. Using phase-space methods we show that this assumption has to be corrected towards the long-wavelength limit. 
 Generalizing the concept of Husimi functions to dielectric interfaces,
 we find that curved interfaces require a semiclassical correction of Fresnel's law due to an interference effect called Goos-H\"anchen shift. It is accompanied by the so-called Fresnel filtering which, in turn, corrects Snell's law. These two contributions are especially important near the critical angle. They are of similar magnitude and correspond to ray displacements in independent phase-space directions that can be incorporated in an adjusted reflection law. 
 We show that deviations from ray-wave correspondence can be straightforwardly understood with 
the resulting adjusted reflection law and discuss its consequences for the phase-space dynamics in optical billiards.
\end{abstract}

\section{Introduction}

Mesoscopic and nanoscopic systems \cite{mesonano}, ranging from quantum dots and nanoparticles to carbon nanotubes and most recently to graphene \cite{graphene_geim}, still receive growing attention. These systems are characterized by a phase-coherence length that is larger than the system size such that interference effects play a crucial role and a quantum mechanical description is in order. Often, however, a semiclassical description is sufficient in order to explain the observations: mesoscopic systems, with typical sizes on the micrometer scale, are truly {\it in the middle} (greek {\it m{\'e}ssi}) between the microscopic and the macroscopic world. The motivation for their investigation comes from both the application-oriented as well as from the basic-research side. Examples are the quest for the miniaturisation of electronic and optical devices or the challenge to observe many-body effects known from bulk metals in finite systems, a prime example here is the observation of the Kondo effect in quantum dots \cite{Kondo_qudot}. Another inspiration comes from the field of quantum chaos \cite{quantumchaos} where the focus lies on the dependence of physical observables on the system geometry. More precisely, the crucial property is whether the dynamics of the underlying classical system is chaotic or integrable, corresponding to a chaotic or integrable (or, in the most generic cases, a mixed) phase space. The first observation of such a sensitivity was in the magnetoconductance through circular (integrable) and stadium-shaped (chaotic) structures that shows a triangular and Lorentzian coherent-backscattering signature, respectively, in the magnetoconductance \cite{chang}.

The objective of the present paper is to give an overview over some of the recent work on quantum chaos and semiclassical aspects in {\it optical} mesoscopic systems which we introduce in the following Section. We will then address a number of deviations from ray-wave correspondence 
and show how they can be explained when correcting the usual specular reflection law for light by semiclassical effects important on the mesoscopic scale. We end the paper with a discussion of the implications of such an adjusted reflection law.

\subsection{Quantum dots and optical microcavities: Billiards for electrons and light}

Quantum dots realized in semiconductor heterostructures are often the first systems that come into mind in the mesoscopic context. Electrons are conveniently trapped and manipulated by various gate electrodes. Especially many-electron ballistic quantum dots are often described as billiards with hard walls and%
used as model systems in the field of quantum chaos \cite{chang}. %

Another class of experimental and theoretical quantum-chaotic model systems are
optical microcavities -- billiards for light instead of electrons. This is possible because of the analogy between the Schr\"odinger and Helmholtz equations for electrons and light, respectively, that holds in two dimensions and for the so-called TM (transverse magnetic) polarization direction (where the magnetic field lies in the resonator plane).
One fundamental difference is, however, the confinement mechanism: For light, there is no charge to manipulate,  and gate-voltage based confinement has to be replaced. The concept used instead is that of total internal reflection at optical interfaces with different refractive indices $n_1$ and $n_2$. Total internal reflection occurs at the optically thinner medium (e.g., when going from glass with refractive index $n_1=n \approx 1.5$ to air with $n_2=1$) for angles of incidence $\chi$ (measured to the boundary normal) larger than the critical angle $\chi_c = \arcsin (1/n)$. The limit $n \to \infty$ corresponds to that of a closed system with hard walls.
This implies in turn that generic optical systems are open systems where the openness is related to, and defined by, the possibility of refractive escape. In other words: Optical microcavities are ideally suited to theoretically as well as experimentally study quantum (wave) chaos in {\it open} systems. Note that the openness is not induced by leads and that we are not interested in transport through the cavity. Rather, the openness exists everywhere along the boundary which is reflected in mixed boundary conditions for the Helmholtz equation with nonvanishing wavefunction 
{\it and} nonvanishing derivative.

\subsection{Ray-wave correspondence!?}
One paradigm intimately related to the field of quantum chaos is the quantum-classical, in the case of optical microcavities, the wave-ray correspondence. It is usually taken for granted and one of the footings of our present understanding of the relation between the classical and the quantum world. We shall see below that, although useful and extremely simple to implement numerically, the ray picture is not able to comprehensively explain {\it all} the results found for optical microresonators. A detailed list of observed deviations will be given in the following section. Note that these deviations are related to the fact that the wavelength $\lambda$ of the electromagnetic field is smaller, but still comparable to the cavity size $R$ (and, therefore, the true ray limit $\lambda \to 0$ is not fulfilled): In typical experimental setups based on semiconductor heterostructures ($n=3.3$), the cavity size is of the order $R \sim 50 {\mu}$m and infrared light with $\lambda \sim 850$ nm is used. In numerical {\it wave} simulations, values $nkR$ exceeding, say, 250, 
are very hard to reach at present.

\begin{figure}[tb]
\begin{center}
\includegraphics[width=11cm]{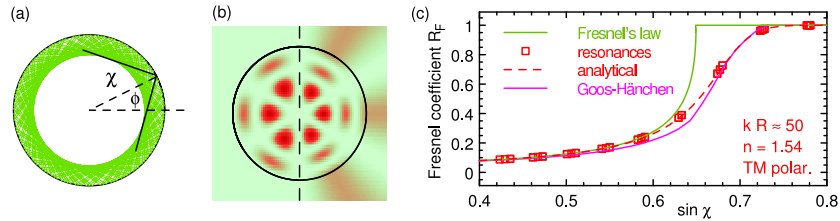}
\end{center}
      \caption{Ray and wave description of the dielectric disk. (a) Whispering-gallery type ray trajectory. (b) Resonances of the closed (left half, corresponding to a wavenumber $kR=9.761$) and dielectric (right half, $n=1.54$, $n k R = 11.428-0.254 i$) disk. The openness of optical systems considerably changes the wave intensity outside the cavity (evanescent wave, or refractive escape). (c) Fresnel's law with deviations from the naive ray-wave correspondence clearly visible in the region of critical incidence. See text for details.}
      \label{fig1}
\end{figure}

\subsection{Description of optical microcavities}

At this point, a few words concerning the description of optical microcavities in the ray and wave picture are in order.

\begin{center}
\begin{figure}[tb]
\includegraphics[width=11cm]{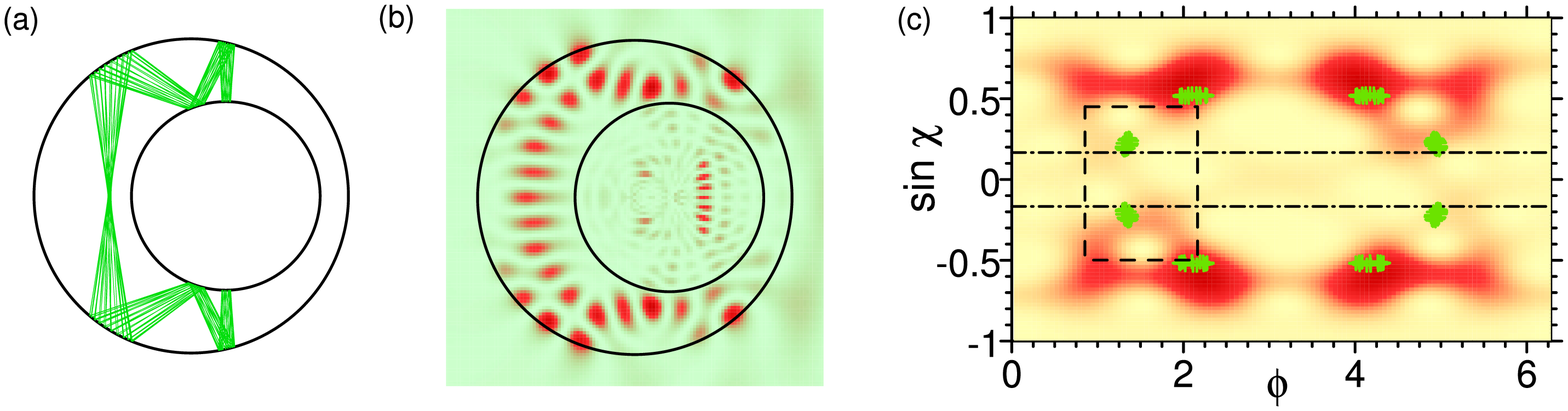}
    \caption{Ray-wave correspondence in the dielectric annular billiard (with refractive indices $n_a=3$ and $n_i=6$ of the annular region and the inner disk). (a) Stable periodic orbit typical for the geometry shown. (b) (One) corresponding resonance wave pattern. (c) Phase-space representation in terms of the Poincar{\'e} surface of section (green/light crosses) and the (inside incident) Husimi function (maxima in red/dark regions) that reveal deviations from ray-wave correspondence on the quantitative level. The dashed rectangle marks a region where violation of forward-backward%
symmetry ($\sin \chi \to - \sin \chi$) is especially evident. The leaky%
region, characterized by refractive escape of rays, is located in between the dashed-dotted lines.}
      \label{fig2}
\end{figure}
\end{center}

{\it Ray picture:} The ray optics description of optical microcavities is based on ray tracing simulations. The trajectory is determined by assuming specular reflections at the resonator walls. 
In addition, one introduces a variable that monitors the (decaying) intensity of the light ray given by Fresnel's law. It is even possible to account for the interference of rays \cite{optlett}. Predictions made for the experimentally accessible far-field intensity 
are now mostly based on the refractive escape of rays taken from the steady-probability distribution \cite{steadyprob}. The%
Poincar{\'e} surface of section summarizes the information about the reflection points at the (outer) resonator boundary in reduced phase space given by a spatial variable parametrizing the position along the boundary (such as the polar angle $\phi$ or the arclength $s$) and the angular momentum
$\sin \chi$,
see Figs.~\ref{fig1}(a) and \ref{fig2}(c). The condition for total internal reflection, $|\sin \chi| > 1/n$, is violated in the so-called leaky (forbidden) region $-1/n < \sin \chi < 1/n$ [marked by dashed-dotted lines in Fig.~\ref{fig2}(c)]. Trajectories hitting this phase-space region will (more or less) easily escape the cavity by refraction.
Periodic orbits with stable islands in the leaky region are considered to be not populated by cavity modes.

{\it Wave Picture:} The objective is to compute the resonances (or quasibound states) of the cavity. The by far most popular approach and a versatile tool is the boundary element method \cite{jan_bem} that gives the resonances directly in the complex plane. The imaginary part of the dimensionless complex wavenumber $k R$ (with $k =  2\pi/\lambda$ and $R$ the radius of curvature) in free space defines the life time and the $Q$-factor of the resonance, $Q=1/[2 \,{\rm Im} (k R)]$. 
An example of a resonant wave pattern is shown in Fig.\ref{fig1}(b) where that of a closed system (left half) is compared with that of an open, optical system with refractive index $n=1.54$ (right half).
Another approach is to use an $S$-matrix method \cite{mhdiss,mhpre} that describes the resonator as an open system being "probed" from outside with (test) plane waves. The position of resonances can be read-off from the Wigner delay time. This approach is straightforwardly implemented%
for (at least partially)
rotationally invariant systems such as the annular billiard, see Fig.~\ref{fig2}(b).

{\it Husimi functions at dielectric interfaces:} 
 The mapping of the resonance wave pattern to phase space is realized by means of the Husimi function \cite{mhhusimi}: Simply speaking, this function measures the overlap of the resonance wave function with a minimal-uncertainty wavepacket centered around a certain position $\phi_0, \chi_0$ in phase space. 
As illustrated for the example of the annular billiard in Fig.~\ref{fig2}, Husimi functions are a particularly useful tool to study ray-wave correspondence. We point out that in order to do so, the concept of Husimi functions has first to be generalized to dielectric interfaces \cite{mhhusimi}.
In hard-wall systems with Dirichlet boundary conditions%
the wave function vanishes at the system's boundary, and the Husimi function can be defined based on the wave function derivative alone. At optical interfaces, however, both the wave function and its derivative are non-zero and it is not clear at all which of the two should be used to define the Husimi function. Moreover, there are now four rays (incident and outgoing on either side of the interface), and the existence of four corresponding Husimi functions would certainly be desirable and support the concept of ray-wave correspondence. As a matter of fact, both issues can (solely) be solved simultaneously as shown in Ref.~\cite{mhhusimi} to which%
we refer the reader for details.

{\it Active (lasing) microcavities:} From an application-oriented point of view, the most interesting application of optical microcavities is that as microlasers. This requires the presence of an active material that allows for lasing operation. Then, in addition to the nonlinearities originating from the (chaotic) resonator geometry those from the lasing operation are important and determine the behaviour. Although a description of the active material can, to a certain degree, be achieved using a complex refractive index $n$ \cite{mhmechprobes}, application of the Schr\"odinger-Bloch model \cite{takaschroedbloch} is in order. We will not further consider the wave description of active cavities in the present paper. We point out that, interestingly, the ray picture (where no active medium can be accounted for) may provide a very reasonable description of the far-field characteristics even for lasing microcavities \cite{tomoko}. Given the goal of building microlasers with directional emission (which comes close to being the holy grail in this field at present), ray simulations have proven to be a valuable tool even away from the ray limit for $\lambda \leq R$ \cite{limacon}.

\section{Deviations from ray-wave correspondence}
\label{sec_deviations}

In the following, we list a number of deviations from ray-wave correspondence that have accumulated over the past, say 10, years. Though each individual fact might be considered just as a slight mismatch, in its entirety these observations suggest that the ray description needs to be corrected away from the exact limit. The necessary corrections will be discussed in the following section, and we finish the paper with an outlook about further consequences of an adjusted ray model.

\subsection{Delayed onset of total internal reflection in Fresnel's law}

One nice example illustrating the interplay between the ray and the wave description is the dielectric disk. On the ray side, the rotational invariance of the system conserves the angle of incidence $\chi$. If $\chi$ is larger than the critical angle, light is confined in a so-called whispering-gallery (WG) orbit, cf.~Fig.~\ref{fig1}(a). On the wave side, this is reflected in an azimuthal quantum number $m$ (e.g., $m=3$ for the WG mode 
in Fig.~\ref{fig1}(b); here, the radial quantum number $\rho =2$) that further characterizes resonances with complex wavenumbers $k R$. There is a one-to-one relation between the ray and wave picture quantities (which are%
$\chi$ and the Fresnel reflection coefficient $R_F (\chi)$ on the ray, and $m$ and $k R$ on the wave side), namely \cite{mhdiss,mhpre} 
\begin{eqnarray}
\sin \chi  & = & \frac{m}{n {\rm Re} (k R) } \:, \\
R_F 	& = & \exp ( 4 n {\rm Im} (k R) \cos \chi) \label{eq_RF} \:. 
	\label{eq:}
\end{eqnarray}
This translation works well, except near the critical angle where the onset of total internal reflection is significantly delayed in the wave picture, see Fig.~\ref{fig1}(c). The green/light (solid) 
curve shows the Fresnel law (for TM polarization) with $R_F=1$ for $\chi \geq \chi_c \approx \arcsin 0.65$. The squares correspond to the Fresnel reflection coefficient, Eq.~(\ref{eq_RF}), for resonances with ${\rm Re} (k R) \approx 50$. Note that a closed analytical expression for $R_F$ (dashed line) was derived and is given in Ref.~\cite{mh_fresnelpre}.

The origin of the deviations near $\chi_c$ must be, and is, the curvature of the dielectric interface.
The question arises (especially when taking a quantum-chaos inspired point of view) whether a semiclassical explanation of this deviation is possible, i.e., one that is based on the ray picture but takes corrections%
originating in the wave character of light into account. That this is indeed possible can be seen by the purple/dark (solid) line for a ray picture completed by such a wave (interference) correction known as the Goos-H\"anchen shift \cite{goos,artmann}
that then closely follows the squares \cite{mh_fresnelpre} .  
We will discuss this effect in detail in Section \ref{sec_ghff}.

\subsection{Correspondence of orbits and resonances in configuration and phase space: Qualitatively, not quantitatively}

In Fig.~\ref{fig2} ray-wave correspondence is illustrated for the example of the dielectric annular billiard. For the geometry chosen, the trajectory/resonance shown in Fig.~\ref{fig2}(a) in the ray, and in Fig.~\ref{fig2}(b) in the wave description belongs to one typical family of stable orbits (or resonances). The similarity between the two patterns is evident and certainly supports the concept of ray-wave correspondence. When going from configuration to phase space, however, the agreement becomes somewhat less convincing, cf. Fig.~\ref{fig2}(c). The larger red (dark) maxima of the (incoming) Husimi function coincide reasonably well with the Poincar\'e signature of the orbit (green crosses), but the smaller maxima (corresponding to the right reflection points closer to the constriction) clearly deviate from the ray signature, which is at least partially related to the Goos-H\"anchen shift mentioned above. Such a behaviour of qualitative, but not quantitative ray-wave correspondence is the typical case. We will investigate the reasons and mechanisms leading to these deviations below.

\subsection{Husimi functions reveal violation of forward-backward%
symmetry}

One intrinsic property of the ray picture is the principle of reversibility of ray trajectories -- in other words, forward-backward or time-reversal symmetry. The corresponding symmetry operation, $\sin \chi \rightarrow -\sin \chi$, is strictly obeyed in the ray picture. This is, however, not the case in the wave description as can be seen in the Husimi function, Fig.~\ref{fig2}(c), especially in the area marked by the dashed rectangle. There are numerous examples of such a behaviour, e.g. \cite{mhpra}, that was recognized in a number of cases but could not be fully understood. Usually, one was content that time-reversal and spatial (mirror axis of the billiard, e.g., $x$-axis in the annular billiard with the symmetry operation $\phi \rightarrow 2 \pi -\phi$) symmetries {\it together} were obeyed. We shall see in the next section that another semiclassical effect, the so-called Fresnel filtering, is responsible for the violation of time-reversal symmetry and implies, when taken into account in an corrected ray picture, non-Hamiltonian dynamics in optical microcavities.

\subsection{Existence of regular modes in chaotic systems}
Optical microresonators studied recently in quite some detail both in theory \cite{koreanerspiral} and in experiment \cite{expspiral}
are cavities with spiral shape, $r(\phi) = R (1+ \epsilon \phi/2 \pi)$, where $\epsilon$ measures the radial offset. In terms of classical dynamics, the spiral billiard is (for all purposes of a physicist) chaotic. The more surprising was the observation of predominantly regular orbits of triangular and star-like shape reported in Ref.~\cite{koreanerspiral}. There are strictly no periodic ray orbits corresponding to these patterns, neither stable (those are missing in chaotic systems) nor unstable (that could become visible as so-called scarred resonances). The observed regular orbits were named quasiscars, and vicinity of the angle of incidence to the critical angle was suspected to be crucial (indeed, triangular orbits were found in cavities with $n=2$, the star-like type for $n=3$).
Note that the existence of these regular orbits suggests that the system apparently possesses rotational symmetry. Clearly, this cannot be the case in a classical description of spiral billiard. The question arises what mechanism re-establishes the rotational invariance in the wave picture. We shall see below that it is again the Fresnel filtering effect that we will explain in detail in Section \ref{sec_ghff}.

\subsection{Violation of Fresnel's and Snell's law at curved interfaces}

\begin{figure}[tb]
\begin{center}
\includegraphics[width=8cm]{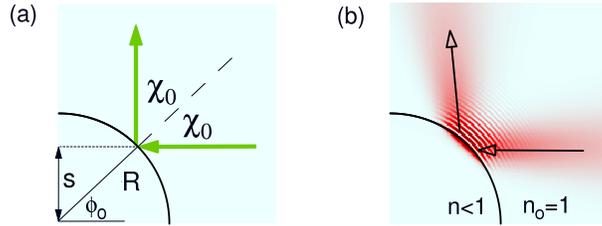}
\end{center}
    \caption{Reflection of (a) a light ray and (b) a Gaussian beam at a curved interface. Clearly, the ray picture prediction fails towards the wave regime $\lambda \leq R$. The reason are Goos-H\"anchen shift and Fresnel filtering, see also Fig.~\ref{fig5}, that vanish only in the pure ray limit $\lambda / R \to 0$.}
      \label{fig3}
\end{figure}

Eventually, we consider the (single) reflection of a Gaussian test ray at a curved interface. For convenience we choose as a cavity an air hole of radius $R$ in a glass matrix, or equivalently, consider a hole with refractive index $n=0.66 <1 $ in air. (Note that this implies concave instead of the convex curvature considered so far.) In Fig.~\ref{fig3} the ray and wave result are shown for the specific case of angle of incidence $\chi_0 = 45^\circ$ (controlled via the impact parameter $s$). Accordingly, the reflected light {\it ray} will leave the cavity vertically. This is, however, not the case in the wave description: Here, the reflected beam leaves the cavity under a larger angle, see Fig.~\ref{fig3}(b).

\section{Correcting ray optics by wave effects: Goos-H\"anchen shift and Fresnel filtering}
\label{sec_ghff}

\begin{figure}[tb]
\begin{center}
\includegraphics[width=11cm]{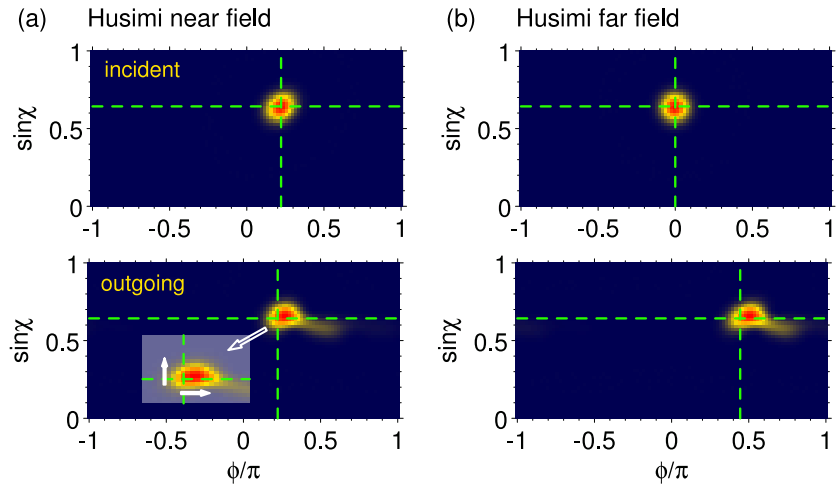}
\end{center}
    \caption{Husimi functions for the reflection of a Gaussian light beam at an circular 
    inclusion with smaller refractive index as shown in Fig.~\ref{fig3}. 
    The parameters of the beam correspond to a light ray incident under $\chi_0 = 40^\circ$, 
    the critical angle is $\chi_c=41.75^\circ$. Shown are the Husimi functions of the incident 
    and reflected beam in the near and in the far field, see text and \cite{mhfresnelprl} for 
    details. The intersection points of the dashed green/light lines indicate the ray model expectations. 	
    Clearly, the signature of the outgoing Husimi function deviates in both phase-space 
    directions from it due to Goos-H\"anchen shift $\Delta \phi_{\rm GH}$ and Fresnel filtering 
    $\Delta \chi_F$.
    }
    \label{fig4}
\end{figure}

At this point a systematic study of the effects causing the deviations from ray-wave correspondence is in order. Inspired by the doubtless advantages of the ray model, such as its easy implementation%
and its conceptual success, 
our objective will be to identify semiclassical corrections%
such that the resulting (adjusted) ray model can quantitatively better capture%
the wave properties of the system.
Following the last example in the previous Section where deviations between the ray and wave behaviour are clearly visible in a single, near-critical reflection, we analyze this situation in some more detail \cite{mhfresnelprl}. Our means of choice are Husimi functions at dielectric interfaces. 
Incident and outgoing Husimi functions for the reflection of a Gaussian beam at a cruved interface, cf.~Fig.~\ref{fig3}, are shown in Fig.~\ref{fig4}.
The intersection of the green/light (dashed) lines marks the position of the maxima as expected from the ray picture.

\begin{figure}[tb]
\begin{center}
\includegraphics[width=9.2cm]{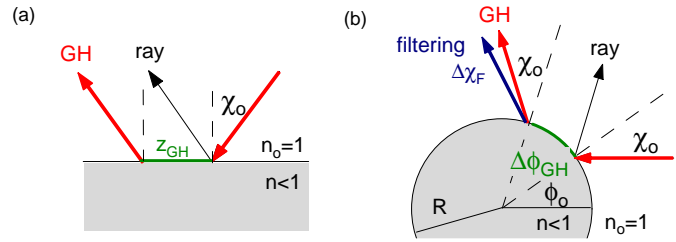}
\end{center}
    \caption{Goos-H\"anchen shift (GH) and Fresnel filtering at (a) a planar and (b) a curved 
    interface. 
    At planar interfaces, a slight variation in the angles of incidence around $\chi_0$ gives rise to the GH shift $z_{\rm GH}$, a lateral shift of the reflected beam in the regime of total internal reflection. At curved interfaces, the GH shift $\Delta \phi_{\rm GH}$ is directly observable as a change in the far field emission direction. In addition, Fresnel filtering $\Delta \chi_F$ leads to non-specular reflection and a deviation from Snell's law that further corrects the far-field emission characteristics. 
    }
    \label{fig5}
\end{figure}

In the near (and similarly in the far) field, the incident Husimi function exactly coincides with the ray picture expectation:%
This choice defines our initial conditions for the incident Gaussian beam (in Fig.~\ref{fig4}, the angle of incidence is $\chi_0 = 40 ^\circ$). 
Deviations become visible when looking at the outgoing Husimi function: The maximum 
deviates from the ray model prediction {\it in two independent directions in phase space}, marked by the arrows in the inset of Fig.~\ref{fig4}: The shift $\Delta \phi_{\rm GH}$ in $\phi$-direction is known as the Goos-H\"anchen shift \cite{goos,artmann}, and the shift $\Delta \chi_F$ in $\chi$-direction has been termed Fresnel filtering \cite{ff}. 
Both effects are schematically illustrated in Fig.~\ref{fig5}(b). Their magnitude depends on the wavenumber and is typically several degrees \cite{mhfresnelprl}, with the Goos-H\"anchen correction being the larger.

Crucial for the understanding of both effects is to realize that in optical microcavities each light ray is actually a light {\it beam}, i.e., composed of light rays with similar but not exactly equal angles of incidence. This becomes immediately clear when recalling that in the mesoscopic regime a light ray assumes a transversal extension of the order $\lambda$.
At a reflection point, the%
corresponding angles of incidence will acquire a certain distribution%
because of the interface curvature \cite{mh_fresnelpre}. In addition, an electromagnetic wave, e.g., a Gaussian beam, always contains a range of angles of incidence. 

Goos and H\"anchen showed in a nice experiment in 1947 \cite{goos} that, for angles of incidence larger than the critical angle, this leads to an interference effect that results in the lateral shift (of the order $\lambda$) of the beam along a {\it planar} interface%
 before it is reflected. This is illustrated in Fig.~\ref{fig5}(a).
The reflection can be thought of as to take place at an effective interface somewhat inside
the optically thinner material.
Whereas the angle of incidence is the {\it same} at the real and the effective interface for {\it planar} interfaces, this is not true at the curved optical boundary of microcavities \cite{mh_fresnelpre}: The angle of incidence is smaller (larger) at convex (concave) boundaries. Applying Fresnel's law to this effective angle of incidence 
can quantitatively explain the deviations in the Fresnel reflections coefficient discussed above, cf.~the purple line in Fig.~\ref{fig1}(c) and Ref.~\cite{mh_fresnelpre} for details.

Fresnel filtering is even more classically to explain than the Goos-H\"anchen shift as there is no underlying interference effect. In a collection of rays with angles of incidence around the critical angle,
the rays with the largest angles will already be totally reflected whereas subcritical rays are still refracted. This can be seen in the signatures of the outgoing Husimi functions in Fig.~\ref{fig4}: The faint signature below the horizontal dashed green/light line corresponds to the refracted beam (that leaves the cavity when hitting the boundary the next time).
It is indeed composed of angles $\chi < \chi_c\approx 41.76^\circ$. In turn, the Husimi signature of the reflected beam is shifted above the%
dashed line that defines the position in the ray model. (Note that the opposite shifts of the reflected and refracted beam ensure conservation of angular momentum.) 
In other words, the maxima of the incident (Gaussian) beam and the reflected (asymmetrically perturbed) beam do not coincide: Around critical incidence, the angle of the reflected beam is always {\it larger} by $\Delta \chi_F$,%
cf.~also Fig.~\ref{fig5}(b), and the law of specular reflection, and consequently forward-backward symmetry, is violated.

Goos-H\"anchen shift and Fresnel filtering have a common origin -- the beam rather than a pure ray nature of electromagnetic {\it waves} -- but act in orthogonal directions in phase space and, therefore, cannot be comprised in one and the same correction to the ray model. On the other hand, they also exhaust the number of possible corrections because there are no more independent directions in phase space. We shall see in the last Section that the different nature of Goos-H\"anchen shift and Fresnel filtering manifests itself in strikingly different effects on the dynamics in optical billiards described with an adjusted ray model that takes the above-discussed non-specular reflection near critical incidence into account \cite{nonhamilt}.

\section{Outlook: Non-Hamiltonian dynamics in quantum-chaotic model systems}

Given the increasing activity in the field of optical microcavities and quantum chaos over the past years, the question arises whether effects of such a non-specular reflection law have not been observed before. We already mentioned that the deviations in Fresnel's law, cf.~Fig.~\ref{fig1}(c), can be fully understood with the Goos-H\"anchen shift. It can also qualitatively explain the differences between ray orbits and resonance patterns, Fig.~\ref{fig2}, via an adjustment at the reflection points 
with the steeper (near critical) angle of incidence. 
In fact, all deviations from ray-wave correspondence discussed in Sec.~\ref{sec_deviations} can be addressed based on Goos-H\"anchen shift (that is important for all angles of incidence $\chi > \chi_c$) and Fresnel filtering (that is important especially for $\chi \approx \chi_c$). The two remaining examples, the existence of regular orbits in spiral microcavities and the breaking of time reversal symmetry in the Husimi functions, can be explained by Fresnel filtering \cite{nonhamilt}. 

In the spiral cavity, the filtering correction $\Delta \chi_F$ in the outgoing beam  re-establishes the conservation of angular momentum as it counteracts the change in curvature that decreases $\chi$. Therefore, regular orbits (similar to those in the disk) may exist again \cite{nonhamilt}.
They are unstable and, consequently, host {\it true} scars.
In general, a finite $\Delta \chi_F$ destroys the principle of ray path reversibility. This is easiest seen when considering a ray with $\chi_0 \approx \chi_c$ that leaves then under an angle $\chi_0 + \Delta \chi_F > \chi_c$. Tracing its trajectory in opposite direction will yield a (nearly) zero filtering correction (such a situation can easily be constructed), and the reflected ray does not coincide with the original ray. It is precisely this type of mechanism that causes the observed loss of time-reversal symmetry in the Husimi functions.

Most remarkably, a billiard dynamics based on the adjusted reflection law leads to non-Hamiltonian dynamics \cite{nonhamilt}. Responsible is the filtering correction that causes deviations of the Jacobian matrix determinant from unity, see \cite{nonhamilt} for details. The origin is the openness of the billiard, effectively described by the adjusted reflection law. Note that dissipative as well as attractive dynamics with the formation of repellors and attractors, respectively, is possible \cite{nonhamilt}. We are optimistic that more signatures of non-Hamiltonian ray dynamics 
will be identified soon%
and that quantum chaos in open systems will remain a fascinating research topic in the future.

%
\vspace*{0.cm}
{\it Acknowledgements.}
The author would like to thank all her co-authors and colleagues for inspiring discussions over the past years.
Sincere thanks go to the Alexander von Humboldt Stiftung (Feodor-Lynen Fellowship 2002-2004) and the Deutsche Forschungsgemeinschaft for ongoing support in the Emmy-Noether Programme and in the Research Group FG 760.

\end{document}